\documentclass[conference]{IEEEtran}
\IEEEoverridecommandlockouts
\usepackage{xcolor}
\usepackage{cite}
\usepackage{amsmath,amssymb,amsfonts}
\usepackage{algorithmic}
\usepackage{graphicx}
\usepackage{textcomp}
\usepackage{xcolor}
\usepackage[graphicx]{realboxes}
\usepackage[figuresleft]{rotating}
\usepackage{url}
\usepackage{tikz}
\usepackage{multirow}
\usetikzlibrary{positioning, fit, arrows.meta, calc, shapes.geometric, backgrounds} 
\usepackage[normalem]{ulem}
\usepackage{booktabs}
\usepackage[caption=false,font=footnotesize]{subfig}
\usepackage{balance}
\usepackage{circuitikz}
\usepackage{etoolbox}

\def\BibTeX{{\rm B\kern-.05em{\sc i\kern-.025em b}\kern-.08em
    T\kern-.1667em\lower.7ex\hbox{E}\kern-.125emX}}

\usepackage[bookmarks=true,hidelinks,pdfpagelabels=true]{hyperref}
\usepackage[printonlyused]{acronym}

\begin{document}

\acrodef{BLL}[BLL]{Bone Length Loss}
\acrodef{CFR}[CFR]{Channel Frequency Response}
\acrodef{CNN}[CNN]{Convolutional Neural Network}
\acrodef{CP}[CP]{cyclic prefix}
\acrodef{CSI}[CSI]{Channel State Information}
\acrodef{CIR}[CIR]{Channel Impulse Response}
\acrodef{DFT}[DFT]{Discrete Fourier Transform}
\acrodef{DL}[DL]{Deep Learning}
\acrodef{FFT}[FFT]{Fast Fourier Transform}
\acrodef{IDFT}[IDFT]{Inverse DFT}
\acrodef{FMCW}[FMCW]{Frequency Modulated Continuous Wave}
\acrodef{OFDM}[OFDM]{Orthogonal Frequency-Division Multiplexing}
\acrodef{ISAC}[ISAC]{Integrated Sensing and Communication}
\acrodef{LS}[LS]{least-squares}
\acrodef{GT}[GT]{Ground-Truth}
\acrodef{GCN}[GCN]{Graph Convolutional Network}
\acrodef{MAE}[MAE]{Masked Autoencoder}
\acrodef{MHSA}[MHSA]{Multi-Head Self-Attention}
\acrodef{MIMO}[MIMO]{Multiple-Input Multiple-Output}
\acrodef{MUSIC}[MUSIC]{Multiple Signal Classification}
\acrodef{MLP}[MLP]{Multilayer Perceptron}
\acrodef{MPJPE}[MPJPE]{Mean Per-Joint Position Error}
\acrodef{MSE}[MSE]{Mean Squared Error}
\acrodef{PA-}[PA-]{Procrustes-aligned}
\acrodef{PDP}[PDP]{Power Delay Profile}
\acrodef{PA-MPJPE}[PA-MPJPE]{Procrustes-aligned MPJPE}
\acrodef{USRP}[USRP]{Universal Software Radio Peripheral}

\title{Human Walking Sensing and Pose Estimation in the 6 GHz Band Using Amplitude and Phase CSI
\thanks{
This work is supported by MSCA TMA DN SMARTTEST. This project has received funding from the European Union’s Horizon Europe - Research and Innovation program - under grant agreement no. 101167834.
}
}

\author{
\IEEEauthorblockN{Zhaorui Yin, Mattia Brambilla, Monica Nicoli}
\IEEEauthorblockA{Politecnico di Milano, Milan, Italy}
}

\maketitle

\begin{abstract}

This paper investigates   human pose estimation from  \ac{OFDM}   signals in an indoor multistatic wireless network operating in the 6\,GHz band.
We design and validate a processing pipeline that exploits both the amplitude and phase of the \ac{CSI} from multiple radio links to estimate the human body pose.
Four deep learning architectures from the literature, namely DT-Pose, MetaFi++, HPE-Li, and VST-Pose, are adapted to the \ac{OFDM} \ac{CSI}  structure and extended to jointly exploit the amplitude and phase information. 
The models estimate the pose of a   human walking within the network coverage area.
Performance evaluation is conducted on an open-access dataset using standard pose-estimation metrics such as Procrustes-aligned Mean Per-
Joint Position Error (PA-MPJPE) and \ac{BLL}.
Results show that reliable human pose reconstruction can be achieved from 6\,GHz \ac{OFDM} \ac{CSI} measurements, with DT-Pose providing the best overall accuracy.
On average, amplitude-only \ac{CSI} yields performance comparable to joint amplitude-phase processing, whereas phase information is more beneficial as a complementary feature rather than as a standalone input.

\end{abstract}
\acresetall

\bstctlcite{BSTcontrol}

\section{Introduction}\label{sec:intro}

Recent advances in wireless sensing have enabled device-free human pose estimation  from \ac{CSI}, offering a non-intrusive alternative to camera-based approaches that can raise privacy concerns \cite{7102722}. 
Applications range from assisted living to safety in industrial environments (e.g., fall detection) and remote healthcare (e.g., elderly after hip-fracture surgery) \cite{savazziDeviceFreeRadioVision2016,6644290,guarino2026survey}. 
Early research in wireless sensing demonstrated that commodity Wi-Fi signals (20~MHz channel in the 2.4~GHz  band)  can be repurposed for device-free sensing \cite{adib2013see}, localizing humans and their gestures behind a wall. 
Capturing the human figure through walls has been demonstrated also with frequencies 5.46--7.24~GHz, with centimeter-level accuracy \cite{adib2015capturing}.
Authors in \cite{10124461} estimated human poses by deriving angle and delay spectra of the multipath signal reflections caused by the human body  in a static environment.
In \cite{JCeS2026}, person identification is framed within a classification problem in \ac{ISAC} for mmWave cellular systems.
Ruan et al.~\cite{ruan2026learning} compare 5G and Wi-Fi \ac{CSI} for indoor localization and show that \ac{CSI} phase information from 5G signals can provide a reliable fingerprint for localization.

A limitation of \ac{CSI}-based pose estimation is its sensitivity to human motion such as walking: a continuous motion typically increases estimation error and challenges simultaneous pose recovery and location tracking~\cite{WiMOSE}. In contrast, prior work suggests that limited bandwidth does not impair \ac{CSI}-based human activity recognition~\cite{cominelli2023exposing}. 
This motivates investigating whether \ac{CSI}-to-pose algorithms can reliably reconstruct the human pose under motion conditions, even with limited-bandwidth signals. 
For this task, \ac{DL} models such as MetaFi++~\cite{zhou2023metafi++}, HPE-Li~\cite{gianHPELiWiFiEnabledLightweight2025}, DT-Pose~\cite{chenRobustRealisticHuman2025} and VST-Pose~\cite{zhangVSTPoseVelocityIntegratedSpatiotemporal2025} have been proposed. 
\ac{CSI}-to-pose models have primarily been developed for  Wi-Fi sensing scenarios, which implicitly constrain key physical-layer parameters such as \ac{CSI} grid resolution, channelization, and signal format.
These assumptions do not necessarily transfer to emerging shared-spectrum scenarios in the 6~GHz license-exempt band ($5.925-7.125$\,GHz), where sensing and communications may rely on custom waveforms and different resource structures.
Indeed, this band has been opened worldwide for license-exempt operation, and both Wi-Fi~6E \cite{80211ax} and NR-U (unlicensed 5G NR) \cite{3gpp_ts37890} 
have been specified to operate therein. Unlicensed rules cover the full spectrum in the United States, whereas in Europe harmonized unlicensed access currently targets only the lower portion ($5.945-6.425$\,GHz).
Wi-Fi and 5G  coexistence and waveform choices are key research study goals to prospective \ac{ISAC} deployments~\cite{coexistance_5gWIFi,9165719}.

To investigate \ac{ISAC} within the 6\,GHz band,  this paper considers the open-access multistatic \ac{OFDM}  dataset in~\cite{usrpofdm}, which  enables the evaluation of device-free human pose estimation using  non-Wi-Fi waveforms in a small indoor environment, aligning with the \ac{ISAC} paradigm. 
We design and validate a processing strategy to estimate the pose of a human moving within the wireless network area. The proposed pipeline incorporates the MetaFi++, HPE-Li, DT-Pose and VST-Pose models, fully exploiting both  amplitude and phase \ac{CSI}. 
A key contribution is the systematic evaluation of \ac{CSI} phase information for human pose estimation. In particular, we compare amplitude-only, phase-only, and joint amplitude-phase representations to assess the contribution of each modality.
Accordingly, the amplitude-based  MetaFi++ and VST-Pose models  are extended to incorporate phase-related features. 
A \ac{LS}  estimation  of the \ac{CIR} is employed to extract the \ac{CSI} from the received signals. The resulting \ac{CSI} tensors serve as input to the \ac{DL} models that estimate the keypoints of skeleton junctions, used to reconstruct the human pose.

\section{System and Dataset}

\definecolor{beaublue}{rgb}{0.74, 0.83, 0.9}

\begin{figure}[!b]
\centering
\resizebox{1\columnwidth}{!}{%
\begin{circuitikz}
\tikzstyle{every node}=[font=\huge]

\def\yTop{14.75}
\def\yBot{11}

\draw [ color={rgb,255:red,255; green,255; blue,255} , fill={rgb,255:red,132; green,192; blue,128}] (17.75,9.5) rectangle (18.75,6.25);

\draw [ fill={rgb,255:red,255; green,255; blue,128} ] (11.25,\yTop) rectangle (16.25,\yBot);

\draw [ fill={rgb,255:red,255; green,255; blue,128} ] (17.25,\yTop) rectangle (22.25,\yBot);

\draw [ fill={rgb,255:red,255; green,255; blue,128} ] (23.25,\yTop) rectangle (28.25,\yBot);

\draw [ fill={rgb,255:red,181; green,181; blue,181} , dashed] (10.25,\yTop) rectangle  (11.25,\yBot);
\draw [ fill={rgb,255:red,192; green,192; blue,192} , dashed] (16.25,\yTop) rectangle  (17.25,\yBot);
\draw [ fill={rgb,255:red,192; green,192; blue,192} , dashed] (22.25,\yTop) rectangle  (23.25,\yBot);

\node at (29.4,12.75) {$\cdots$};

\draw [ fill={rgb,255:red,192; green,192; blue,192} , dashed] (30.25,\yTop) rectangle  (31.25,\yBot);

\draw [dashed] (10.25,16) -- (10.25,\yTop);

\draw [<->, >=Stealth] (10.25,16) -- (16.25,16)
  node[midway, fill=white] {$2+N_\mathrm{s}$};

\node [font=\LARGE, rotate around={90:(0,0)}] at (22.75,13) {sync};
\node [font=\LARGE, rotate around={90:(0,0)}] at (10.75,13) {sync};
\node [font=\LARGE, rotate around={90:(0,0)}] at (16.75,13) {sync};
\node [font=\LARGE, rotate around={90:(0,0)}] at (30.75,13) {sync};

\draw [short] (12,11) -- (15,9.75);
\draw [short] (11.25,11) -- (10,9.75);

\draw [dashed] (11.25,15.25) -- (11.25,\yTop);
\draw [dashed] (16.25,16) -- (16.25,\yTop);
\draw [dashed] (22.25,16) -- (22.25,\yTop);
\draw [dashed] (28.25,16) -- (28.25,\yTop);

\node [font=\LARGE] at (10.5,16.5) {$t = 0$};
\node [font=\LARGE] at (16.25,16.5) {$t = 1$};
\node [font=\LARGE] at (22.25,16.5) {$t = 2$};
\node [font=\LARGE] at (28.5,16.5) {$t = 3$};

\draw [short] (12,11) -- (12,\yTop);

\node [font=\LARGE] at (12.5,9.5) {$1^{\mathrm{st}}$ symbol};

\draw [short] (10,9.75) -- (10,9);
\draw [short] (15,9.75) -- (15,9.25);

\draw [<->, >=Stealth] (11.25,7.5) -- (15,7.5)
  node[midway, fill=white] {$N_\mathrm{DFT}$};

\draw [<->, >=Stealth] (11.25,15.25) -- (16.25,15.25)
  node[midway, fill=white] {$N_\mathrm{s}$};

\draw [->, >=Stealth] (15,8.5) -- (17.5,8.5);
\node [font=\LARGE] at (16.25,9) {DFT};

\draw [short] (12.75,\yTop) -- (12.75,\yBot);
\draw [short] (13.5,\yTop)  -- (13.5,\yBot);
\draw [fill=white, draw=none] (14.2,14.75-0.01) rectangle (15.45,11.0+0.01);
\draw [short] (14.2,\yTop)  -- (14.2,\yBot);
\draw [short] (15.5,\yTop)  -- (15.5,\yBot);
\node at (14.95,12.8) {$\cdots$};
\node [rotate=90,font=\LARGE] at (13.8,12.85) {$\mathbf{y}_{\ell,1,i}$};


\draw [short] (18.0,\yTop) -- (18.0,\yBot);
\draw [short] (18.75,\yTop) -- (18.75,\yBot);
\draw [short] (19.5,\yTop)  -- (19.5,\yBot);
\draw [fill=white, draw=none] (20.2,14.75-0.01) rectangle (21.45,11.0+0.01);
\draw [short] (20.2,\yTop)  -- (20.2,\yBot);
\draw [short] (21.5,\yTop)  -- (21.5,\yBot);
\node at (20.95,12.8) {$\cdots$};
\node [rotate=90,font=\LARGE] at (19.8,12.85) {$\mathbf{y}_{\ell,2,i}$};

\draw [short] (24.0,\yTop) -- (24.0,\yBot);
\draw [short] (24.75,\yTop) -- (24.75,\yBot);
\draw [short] (25.5,\yTop)  -- (25.5,\yBot);
\draw [fill=white, draw=none] (26.2,14.75-0.01) rectangle (27.45,11.0+0.01);
\draw [short] (26.2,\yTop)  -- (26.2,\yBot);
\draw [short] (27.5,\yTop)  -- (27.5,\yBot);
\node at (26.95,12.8) {$\cdots$};
\node [rotate=90,font=\LARGE] at (25.8,12.85) {$\mathbf{y}_{\ell,3,i}$};

\draw [ fill={rgb,255:red,255; green,255; blue,128} ] (11.25,9) rectangle  node {\LARGE $\mathbf{y}_{\ell,1,1}$} (15,8);

\draw [fill={rgb,255:red,192; green,192; blue,192}, dashed] (10,9) rectangle node {\LARGE CP} (11.25,8);
\draw [short] (15,9.75) -- (15,9);

\draw [ fill={rgb,255:red,192; green,192; blue,192} , dashed] (17.75,9.75) rectangle  (18.75,9.5);
\draw [ fill={rgb,255:red,192; green,192; blue,192} , dashed] (17.75,6.25) rectangle  (18.75,6);
\node [font=\LARGE] at (16,6) {guard band};

\node [font=\LARGE, rotate around={90:(0,0)}] at (18.25,8) {$\mathbf{Y}_{\ell,1,1}$};

\draw [ color={rgb,255:red,255; green,255; blue,255} , fill={rgb,255:red,132; green,192; blue,128}] (18.75,9.5) rectangle (19.75,6.25);
\draw [ fill={rgb,255:red,192; green,192; blue,192} , dashed] (18.75,9.75) rectangle  (19.75,9.5);
\draw [ fill={rgb,255:red,192; green,192; blue,192} , dashed] (18.75,6.25) rectangle  (19.75,6);
\node [font=\LARGE, rotate around={90:(0,0)}] at (19.25,8) {$\mathbf{Y}_{\ell,1,2}$};

\draw [ color={rgb,255:red,255; green,255; blue,255} , fill={rgb,255:red,132; green,192; blue,128}] (19.75,9.5) rectangle (20.75,6.25);
\draw [ fill={rgb,255:red,192; green,192; blue,192} , dashed] (19.75,9.75) rectangle  (20.75,9.5);
\draw [ fill={rgb,255:red,192; green,192; blue,192} , dashed] (19.75,6.25) rectangle  (20.75,6);
\node [font=\LARGE, rotate around={90:(0,0)}] at (20.25,8) {$\mathbf{Y}_{\ell,1,3}$};

\draw [ color={rgb,255:red,255; green,255; blue,255} , fill={rgb,255:red,132; green,192; blue,128}] (22.25,9.5) rectangle (23.25,6.25);
\draw [ fill={rgb,255:red,192; green,192; blue,192} , dashed] (22.25,9.75) rectangle  (23.25,9.5);
\draw [ fill={rgb,255:red,192; green,192; blue,192} , dashed] (22.25,6.25) rectangle  (23.25,6);
\node [font=\LARGE, rotate around={90:(0,0)}] at (22.75,8) {$\mathbf{Y}_{\ell,1,N_\mathrm{s}}$};

\draw [->, >=Stealth] (11.75,11) .. controls (18.75,9.5) and (18.25,10.5) .. (18.25,9.75) ;
\draw [->, >=Stealth] (12.5,11) .. controls (19.75,9.5) and (19.25,10.5) .. (19.25,9.75) ;
\draw [->, >=Stealth] (13.25,11) .. controls (19.5,9.75) and (20.25,10.5) .. (20.25,9.75) ;
\draw [->, >=Stealth] (16,11) .. controls (22.25,9.75) and (22.75,10.75) .. (22.75,9.75) ;

\node at (21.65,8) {$\cdots$};

\draw [<->, >=Stealth] (23.75,9.5) -- (23.75,6.25)
  node[midway, fill=white] {$K$};

\draw [dashed] (23.25,9.5) -- (31.25,9.5);
\draw [dashed] (23.25,6.25) -- (31,6.25);

\draw [->, >=Stealth] (24.5,7.75) -- (29.25,7.75);
\node [font=\LARGE] at (26.75,8.25) {LS-estimate};

\def\hL{29.75}
\def\hR{34.75}
\def\hT{9.5}
\def\hB{6.25}

\draw[fill=beaublue] (\hL,\hT) rectangle (\hR,\hB);
\draw[short] ({\hL+0.75+0.1},\hT) -- ({\hL+0.75+0.1},\hB);
\draw[short] ({\hL+1.50+0.3},\hT) -- ({\hL+1.50+0.3},\hB);
\draw[short] ({\hL+2.25+0.5},\hT) -- ({\hL+2.25+0.5},\hB);

\draw[fill=white, draw=none] ({\hL+2.95-0.17},\hT-0.01) rectangle ({\hL+4.20-0.15},\hB+0.01);

\draw[short] ({\hL+4.25-0.2},\hT) -- ({\hL+4.25-0.2},\hB);

\pgfmathsetmacro{\hMid}{(\hT+\hB)/2}
\node at ({\hL+3.60-0.2},{\hMid}) {$\cdots$};
\node[rotate=90,font=\LARGE] at ({\hL+0.375+0.05},\hMid) {$\widehat{\mathbf{H}}_{\ell,1,i}$};
\node[rotate=90,font=\LARGE] at ({\hL+1.125+0.25},\hMid) {$\widehat{\mathbf{H}}_{\ell,2,i}$};
\node[rotate=90,font=\LARGE] at ({\hL+1.875+0.45},\hMid) {$\widehat{\mathbf{H}}_{\ell,3,i}$};
\node[rotate=90,font=\LARGE] at ({\hR-0.375-0.1},\hMid) {$\widehat{\mathbf{H}}_{\ell,N_\mathrm{s},i}$};

\draw [ fill={rgb,255:red,255; green,255; blue,128} ] (31.25,\yTop) rectangle (36.25,\yBot);

\draw [short] (32,\yTop) -- (32,\yBot);
\draw [short] (32.75,\yTop) -- (32.75,\yBot);
\draw [short] (33.5,\yTop)  -- (33.5,\yBot);
\draw [fill=white, draw=none] (34.2,14.75-0.01) rectangle (35.45,11.0+0.01);
\draw [short] (34.2,\yTop)  -- (34.2,\yBot);
\draw [short] (35.5,\yTop)  -- (35.5,\yBot);
\node at (34.95,12.8) {$\cdots$};
\node [rotate=90,font=\LARGE] at (33.8,12.85) {$\mathbf{y}_{\ell,N_\mathrm{f},i}$};

\draw [dashed] (36.25,16) -- (36.25,\yTop);
\node [font=\LARGE] at (36.5,16.5) {$t = N_\mathrm{f}$};

\draw [<->, >=Stealth] (35.25,9.5) -- (35.25,6.25)
  node[midway, fill=white] {$K$};

\draw [<->, >=Stealth] (29.75,10) -- (34.75,10)
  node[midway, fill=white] {$N_\mathrm{s}$};



\end{circuitikz}
}%
\caption{Frame structure and processing of the received \ac{OFDM} signal for \ac{CSI} estimation.
}
\label{fig:my_label}
\end{figure}
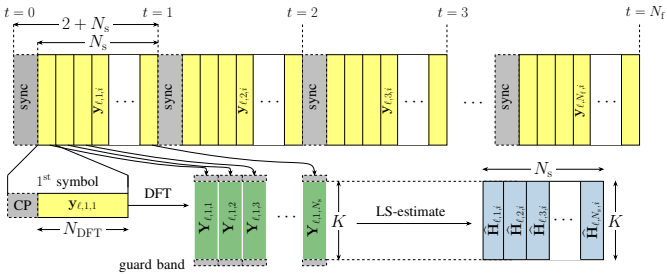

\subsection{Radio Sensing Scenario and CSI Estimation}
We consider a multistatic radio-sensing scenario where a set of radio devices transmit a continuous \ac{OFDM} data stream. The transmission is organized into frames, each preceded by a short synchronization (sync) field used for frame alignment at the receiver side. 
Wireless signals are used to sense the activity of a human body moving in the coverage area of the local wireless network.
The  communication links are indexed by $\ell \in \{1,\ldots,L\}$.  
Referring to Fig.~\ref{fig:my_label}, the received \ac{OFDM} signal on link $\ell$ at time frame $t = 1,\ldots,N_\mathrm{f}$ is $\mathbf{y}_{\ell,t,i}\in\mathbb{C}^{N_\mathrm{DFT}\times 1}$, where index $i = 1,\ldots,N_\mathrm{s}$ refers to the  $i$-th symbol out of $N_\mathrm{s}$ per frame. $N_\mathrm{DFT}$ is the \ac{DFT} size of the \ac{OFDM} symbol. Out of $ N_\mathrm{DFT}$ subcarriers,  $K = N_\mathrm{DFT} - N_\mathrm{guard}$ are usable while $N_\mathrm{guard}$ are guard symbols. The usable subcarriers are indexed as $k\in \{N_\mathrm{guard}/2,\ldots,(N_\mathrm{DFT} - N_\mathrm{guard}/2 - 1)\}$. By applying the \ac{DFT} we obtain the frequency representation $\mathbf{Y}_{\ell,t,i}\in\mathbb{C}^{K\times 1}$.

Being $\mathbf{X}_{\ell,t,i}\in\mathbb{C}^{K\times 1}$ the transmit data  
on the $K$ subcarriers for link $\ell$, frame $t$ and symbol $i$,
the received signal is:
\begin{align}
\mathbf{Y}_{\ell,t,i} &=
\mathrm{diag}\!\left(\mathbf{X}_{\ell,t,i}\right) \mathbf{H}_{\ell,t,i} + \mathbf{N}_{\ell,t,i} ,
\label{eq:ofdm_mimo_fd}
\end{align}
where $\mathbf{H}_{\ell,t,i} \in\mathbb{C}^{K\times 1}$ is the \ac{CFR}
and $\mathbf{N}_{\ell,t,i}\in\mathbb{C}^{K\times 1}$ is the background noise including the interference. 
The frequency response is related to the time-domain \ac{CIR}
$\mathbf{h}_{\ell,t,i}\in\mathbb{C}^{N_\tau\times 1}$ via a partial \ac{DFT}  as $\mathbf{H}_{\ell,t,i} = \mathbf{F}\,\mathbf{h}_{\ell,t,i}$.
The \ac{DFT} matrix   $\mathbf{F}\in\mathbb{C}^{K\times N_\tau}$
has entries
$[\mathbf{F}]_{k,\tau}=\frac{1}{\sqrt{K}}\exp\!\left(-j\frac{2\pi}{K}k\tau\right)$ for $\tau=\{0,\ldots,N_\tau-1\}$, where $\tau$ and $N_\tau$ denote the delay index and the  maximum temporal support of the channel, assumed to be within the \ac{CP}. 
Assuming the matrix $\mathbf{B}_{\ell,t,i} = \mathrm{diag}\!\left(\mathbf{X}_{\ell,t,i}\right)\mathbf{F} $
as full column rank, the \ac{LS} estimate of the per-link \ac{CSI} is:
\begin{equation}
\widehat{\mathbf{H}}_{\ell,t,i} =
\mathbf{F}
\underbrace{
\mathbf{B}_{\ell,t,i}^{\mathsf{\dagger}}
\mathbf{Y}_{\ell,t,i}}_{\widehat{\mathbf{h}}_{\ell,t,i}}
= \mathbf{H}_{\ell,t,i} +
\underbrace{\mathbf{F}\mathbf{B}_{\ell,t,i}^{\mathsf{\dagger}} \mathbf{N}_{\ell,t,i}}
_{\widetilde{\mathbf{N}}_{\ell,t,i} } ,
\label{eq:fullk_cfr_recon}
\end{equation}
where $\mathbf{B}_{\ell,t,i}^{\mathsf{\dagger}} = (\mathbf{B}_{\ell,t,i}^{\mathsf{H}}\mathbf{B}_{\ell,t,i})^{-1}\mathbf{B}_{\ell,t,i}^{\mathsf{H}}$ and $\widetilde{\mathbf{N}}_{\ell,t,i}$ accounts for the measurement noise.
Stacking the $N_{\mathrm{s}}$ symbols of the $t-$th frame, we obtain $\widehat{\mathbf{H}}_{\ell,t} = [ \widehat{\mathbf{H}}_{\ell,t,1}^\top \cdots \widehat{\mathbf{H}}_{\ell,t,N_{\mathrm{s}}}^\top]^\top$, which can be seen as the combination of a static multipath component originated from the environment 
and a dynamic human-related multipath component 
from the target to be sensed.

\subsection{Dataset Description and Visualization}

\input{FigTex/figure2}

We refer to the dataset in~\cite{usrpofdm} for designing and validating human walking sensing and pose estimation. It has been collected in an area of $A=3\times 3 \,\mathrm{m}^2$ with human walking at a speed of about $1$\,m$/$s for $10$ different rounds.  Three \acp{USRP} X410  operate at carrier frequencies $6.74$, $6.80$, and $6.86$~GHz in a $3\times 3$ multistatic configuration, where each \ac{USRP} transmits on a single frequency but receives on all frequencies (see top area of Fig.~\ref{fig:system_architecture}). 
Each transmitter emits \ac{OFDM} signals with $16$~MHz bandwidth, $N_{\mathrm{DFT}}=2048$, and cyclic prefix length of $512$ samples.
At the receiver, in the baseband, after discarding the cyclic prefix and removing $128$ guard band subcarriers, we retain $1920$ subcarriers and $100$ \ac{OFDM} symbols for \ac{CSI} estimation at $20$\,MHz sampling rate. 
The parameters of the dataset~\cite{usrpofdm} are reported in Table~\ref{tab:sys_params_usrpofdm}.

\begin{table}[t]
\centering
\caption{System parameters of the multistatic \ac{OFDM} dataset~\cite{usrpofdm}.}
\label{tab:sys_params_usrpofdm}
\vspace{-7pt}
\begin{tabular}{l c}
\toprule
System parameter & Value \\
\midrule
Carrier frequencies & $\{6.74\, \,\mathrm{/} \,\,6.80\,\, \mathrm{/}\,\,6.86\}~\mathrm{GHz}$ \\
Number of multistatic links & $9$ \\
Sampling rate & $20\, \mathrm{MHz}$ \\
Signal bandwidth & $16~\mathrm{MHz}$ \\
OFDM subcarriers & $2048$ \\
Guard subcarriers & $128$  \\
Cyclic prefix length & $512$ samples \\
Symbols per frame & $100$ \\
Time per frame & $12.8\,\mathrm{ms}$ \\
Number of human keypoints (extrap.) & $8\,\,(13)$ \\
Walking area & $3\,\mathrm{x}\, 3\,\mathrm{m}^2$ \\
\bottomrule
\end{tabular}
\end{table}

\begin{figure}
    \centering

    \subfloat[Human walking scenario \label{fig:human_walking_scenario}]{%
       \includegraphics[width=0.49\columnwidth]{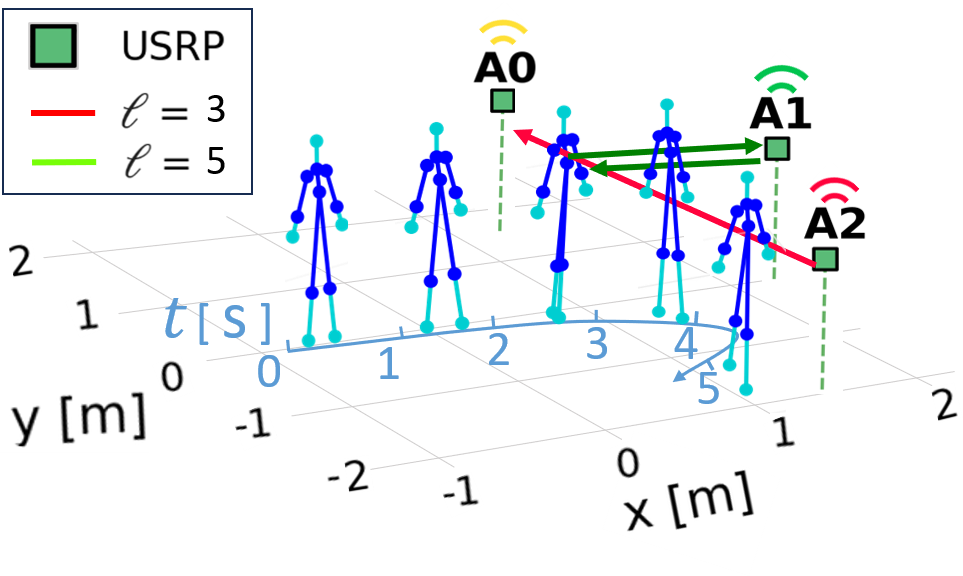}
       }
      \subfloat[\ac{PDP}, $\ell = 3$ \label{fig:PDF_link3}]{%
       \includegraphics[width=0.49\columnwidth]{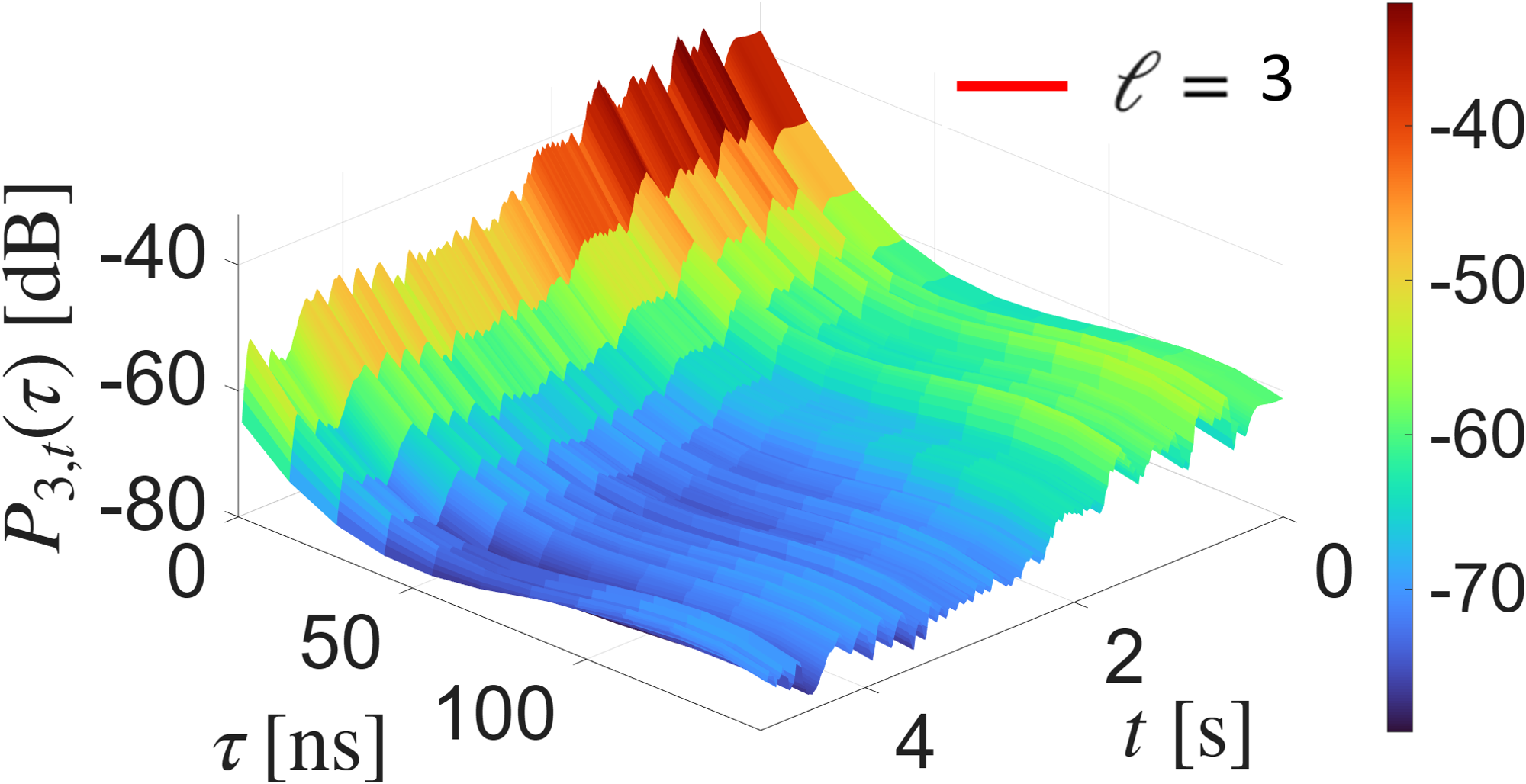}
       }

       \subfloat[\ac{PDP}, $\ell = 5$\label{fig:PDF_link5}]{%
       \includegraphics[width=0.49\columnwidth]{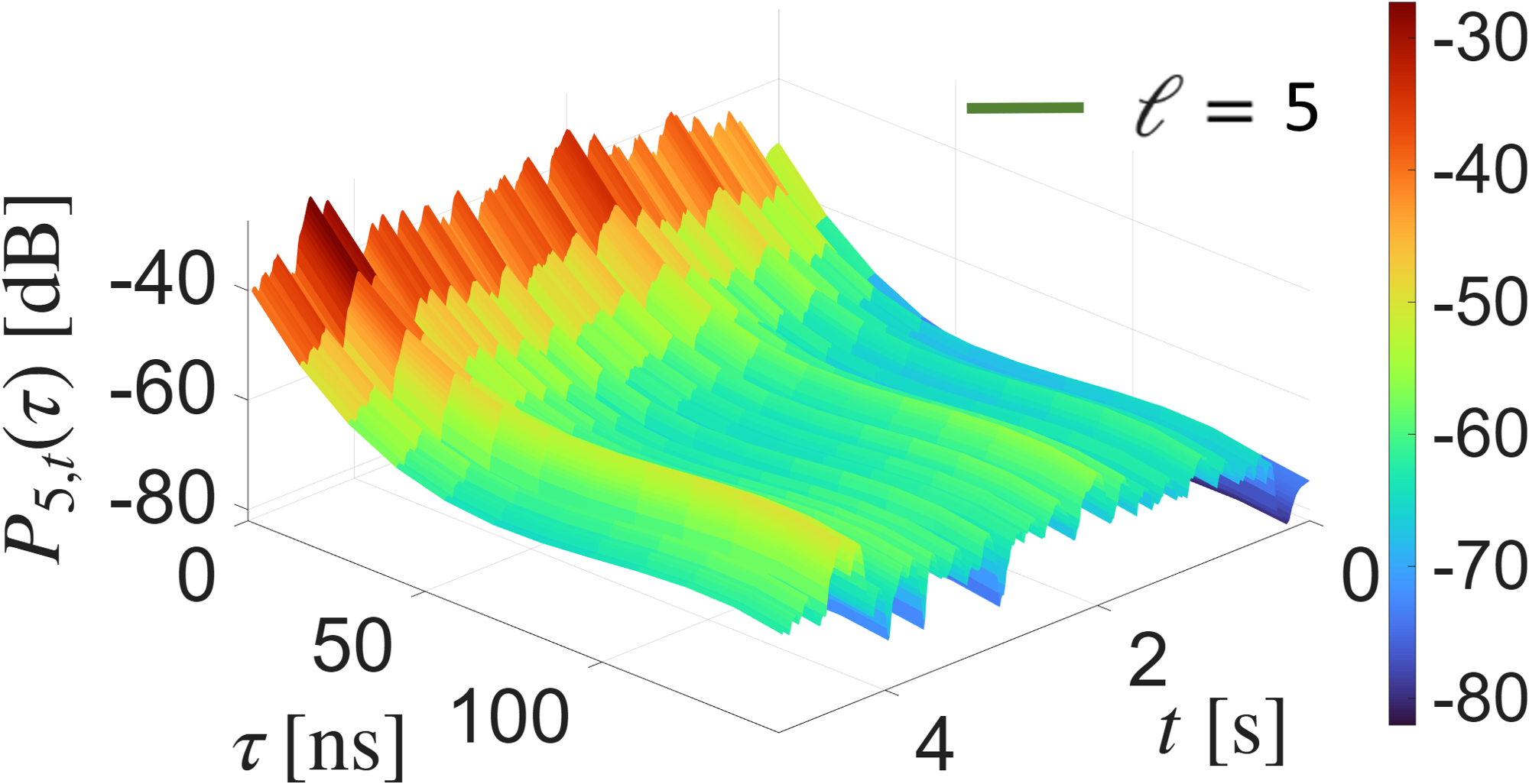}
       }
               \hfill
        \subfloat[Phase, $\tau = 0$\label{fig:phase_evolution}]{%
        \includegraphics[width=0.47\columnwidth]{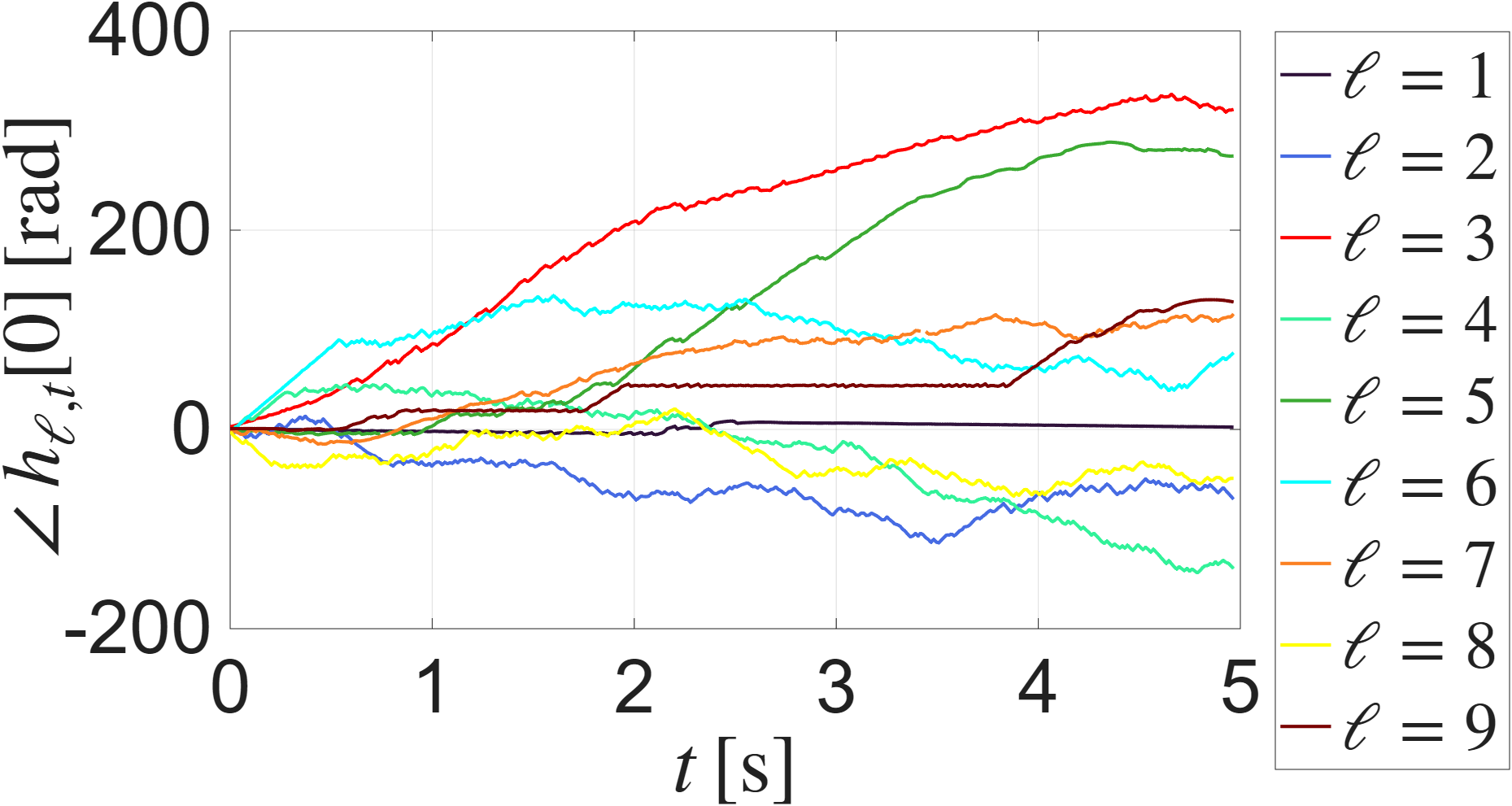}
        }

    \caption{(a) Human walking captured from $t=0\,$s to $t=5$\,s by three \acp{USRP} (A0, A1 and A2). (b) \ac{PDP} for link $\ell = 3$ (bistatic). (c) \ac{PDP} for link $\ell = 5$ (monostatic). (d) Phase evolution of the first tap ($\tau=0$) for all links.
}
\vspace{-8pt}
    \label{fig:pdp}
\end{figure}

The dataset is illustrated in Fig. \ref{fig:human_walking_scenario}, which shows the human motion over $5$\,s in the area of three \acp{USRP}, and highlights two wireless links: $\ell=3$ (bistatic at $f_c=6.74$\,GHz) and $\ell=5$ (monostatic at $f_c=6.80$\,GHz). 
The evolution of the channel \ac{PDP} for these two links is shown in Fig. \ref{fig:PDF_link3} and Fig. \ref{fig:PDF_link5}. The \ac{PDP} is computed for delay $\tau$ as:
\begin{align}
P_{\ell,t}(\tau)
=\frac{1}{N_s}\sum_{i=1}^{N_s}\left|\widehat{h}_{\ell,t,i}[\tau]\right|^{2},
\label{eq:pdp_scalar}
\end{align}
where $\widehat{h}_{\ell,t,i}[\tau]$ denotes the $\tau$-th tap of the estimated \ac{CIR} vector $\widehat{\mathbf{h}}_{\ell,t,i}$.
The figures highlight that the human body perturbs the received signal as it moves closer to the radios.  Accordingly, the \ac{PDP}  is non-stationary: the bistatic link 
(Fig.~\ref{fig:PDF_link3}) shows a power reduction when the human approaches the link (from $t=2$ to $t=4$) while for the monostatic link (Fig.~\ref{fig:PDF_link5}) the power increases with time thanks to the human reflection.
Therefore, we examine the channel evolution over time $t$ and across all  links $L$ to capture the multipath dynamics that are informative for human pose reconstruction. Besides, we compare the phase components across all links (see Fig.~\ref{fig:phase_evolution}) to showcase a difference over $t$. As a result, the \ac{CIR} becomes time-varying, and the corresponding \ac{PDP} ${P}_{\ell,t}$ exhibits noticeable temporal dynamics. Overall, these amplitude- and phase-dependent variations are jointly represented in the estimated \ac{CSI} tensor $\widehat{\mathbf{H}}_{\ell,t}$ and are exploited for the final human pose estimation by the algorithms described in the next section.

\section{CSI-to-Pose DL Models}

In this section, we propose a methodology to estimate the human pose exploiting the whole \ac{CSI} tensor in amplitude and phase of all $L$ links
$\bigl\{\,|\widehat{\mathbf{H}}_{\ell,t}|,\,\angle\widehat{\mathbf{H}}_{\ell,t}\,\bigr\}^L_{\ell=1}$,
which are used as inputs to the \ac{DL} models. 
The dataset  uses carrier frequencies $f_c$ at $6.74$, $6.80$ and $6.86$ GHz, resulting in a wavelength in the order of $\lambda = c/f_c$ = 4.4\,cm.
This motivates the evaluation of phase as an additional feature that may provide complementary information for human pose estimation. 
While the delay resolution is mainly determined by the signal bandwidth, the wavelength affects the resolution of phase-based sensing, as it defines the spatial scale at which small body motions can induce measurable perturbations in the \ac{CSI} \cite{personinwifi}. 
In fact, amplitude-only \ac{CSI} sensing is constrained by the limited delay resolution associated with the signal bandwidth, 
whereas phase \ac{CSI} measurements provide higher resolution.

\subsection{Model Description}

A processing pipeline is designed to infer the human pose from $\bigl\{|\widehat{\mathbf{H}}_{\ell,t}|,\,\angle\widehat{\mathbf{H}}_{\ell,t}\bigr\}^L_{\ell=1}$.
It starts by removing the guard bands $N_\mathrm{guard}$ and grouping \ac{OFDM} symbols, then collecting the \ac{CSI} information on amplitude and phase. The amplitude component is processed with a discrete wavelet filtering to suppress high-frequency noise~\cite{zhangVSTPoseVelocityIntegratedSpatiotemporal2025} while the phase undergoes an unwrapping operation and PhaseFi denoising~\cite{PhaseFi}. 
A 2-in-1 \ac{MLP} with four hidden neurons is used to fuse the amplitude and phase features before feeding them into the \ac{DL} backbone, such that the input tensor has dimension $ L \times K \times N_{\text{s}} \, (9\times 1920 \times 100)$.
Then, four \ac{DL} architectures for pose estimation are introduced in the processing chain and used for the final prediction: MetaFi++~\cite{zhou2023metafi++}, HPE-Li~\cite{gianHPELiWiFiEnabledLightweight2025}, DT-Pose~\cite{chenRobustRealisticHuman2025} and VST-Pose~\cite{zhangVSTPoseVelocityIntegratedSpatiotemporal2025}. 
Since these models were originally proposed for Wi-Fi systems, they required an input tensor of dimension $114 \times 10$, with $114$ subcarriers and $10$ symbols in a \ac{CSI} tensor. Here, we modify the input \ac{CNN} and \ac{MAE} layer to support the input dimension with $K=1920$ subcarriers and $N_\mathrm{s} = 100$ symbols. 

MetaFi++~\cite{zhou2023metafi++} follows a \ac{CNN}-transformer design, applying a ResNet-based \ac{CNN} to extract a compact \ac{CSI} feature map and flattening it to \ac{MHSA}-based integration. 
DT-Pose~\cite{chenRobustRealisticHuman2025} uses \ac{MAE}-style self-supervised pretraining and a topology-constrained decoder that combines \ac{MAE} and \ac{GCN} based adjacent-joint aggregation with transformer-based joint modeling. 
HPE-Li~\cite{gianHPELiWiFiEnabledLightweight2025} replaces an explicit transformer with dual-domain attention implemented in convolution via DSKConv, learning kernel selection over \ac{CSI} features.
VST-Pose~\cite{zhangVSTPoseVelocityIntegratedSpatiotemporal2025} employs a dual-stream spatiotemporal transformer with an explicit velocity branch, applying temporal and spatial \ac{MHSA} to joint-wise embeddings.

\subsection{Model Output and Loss Function}
After the \ac{DL} networks, a linear pose decoder extracts $J$ anatomical landmarks, i.e. 3D keypoints,
denoted as $\widehat{\mathbf{p}}_{j}^{(t)}$, $j\in \{1, \ldots, J \}$, representing the estimate of the \ac{GT} keypoints ${\mathbf{p}}_{j}^{(t)}\in\mathbb{R}^{3}$.
In the dataset~\cite{usrpofdm}, a motion-capture system with frame rate of $100$\,fps is used to get the \ac{GT} for $8$ human joints at chests, elbows and knees. 
To obtain a standard human representation, we augment the \ac{GT} points to $J=13$ by geometrically performing linear extrapolation from the elbow and knee joints to synthesize hand and foot keypoints from nearby ones. Specifically, replacing the subscript $j$ with the associated physical body part, and considering time indexed over $t$, we obtain the hand and foot keypoints as:
\begin{align}
    {\mathbf{p}}^{(t)}_{\text{hand}} 
    &=
    \mathbf{p}^{(t)}_{\text{elbow}} + \label{eq:hand}
    \tfrac{1}{2}(\underbrace{\mathbf{p}^{(t)}_{\text{elbow}}-\mathbf{p}^{(t)}_{\text{shoulder}}}_{\mathrm{arm\,\,extend}})
    + \underbrace{\mathbf{p}^{(t)}_{\text{elbow}} - \mathbf{p}^{(t-1)}_{\text{elbow}}}_{\mathrm{arm\,swing}} ,
\\
    {\mathbf{p}}^{(t)}_{\text{foot}}
    &=
    \mathbf{p}^{(t)}_{\text{knee}} + \label{eq:foot}
    \tfrac{1}{2}(\underbrace{\mathbf{p}^{(t)}_{\text{knee}}-\mathbf{p}^{(t)}_{\text{chest}}}_{\mathrm{leg\,\,extend}})
    + \underbrace{\mathbf{p}^{(t)}_{\text{knee}} - \mathbf{p}^{(t-1)}_{\text{knee}}}_{\mathrm{leg\,swing}}.
\end{align}

The extrapolation  in \eqref{eq:hand} and \eqref{eq:foot} is velocity-based along the walking direction, where the velocity is estimated from the relative motion between the elbow and knee to approximate arm and leg swinging motion beyond their joints. 
In addition, we reconstruct the head keypoint by extending upward from the mid-point of the chest by a fixed head–neck length of 20\,cm, yielding a more complete skeletal representation. The resulting effect is illustrated in Fig.~\ref{fig:system_architecture} (bottom right), where blue keypoints indicate the original \ac{GT} keypoints as in \cite{usrpofdm} while the extrapolated ones are in light blue. The height of the human body after extrapolation is approximately 160\,cm.

The models are trained using the \ac{MSE} loss function, averaged over the keypoints and frames:
\begin{equation}
\mathcal{L}_{\mathrm{MSE}}
= \frac{1}{N_\mathrm{f}\,J}\sum_{t=1}^{N_\mathrm{f}}\sum_{j=1}^{J}
\left\lVert \widehat{\mathbf{p}}_{j}^{(t)} - \mathbf{p}_{j}^{(t)} \right\rVert_2^2.
\label{eq:mse_pose}
\end{equation}
The same loss is also used as a validation metric.
All models are optimized using Adam optimizer with learning rate $10^{-4}$ and trained on a workstation equipped with an NVIDIA RTX A5000 GPU with 24\,GB memory. 
\section{Evaluation of Human Pose Estimation}

\subsection{Performance Metrics}
For evaluation, we also report the
\ac{PA-MPJPE} metric, defined as the \ac{MPJPE}~\cite{9320400} after aligning each predicted pose to the reference by a similarity transform with scaling $s^{(t)}$, rotation $\mathbf{R}^{(t)}$, translation $\mathbf{v}^{(t)}$.  The \ac{PA-MPJPE} 
is defined as:
\begin{equation}
\mathrm{PA\text{-}MPJPE}
= 
\frac{1}{N_\mathrm{f}J}\sum_{t=1}^{N_\mathrm{f}}
\sum_{j=1}^{J}
\left\lVert
s^{(t)}\mathbf{R}^{(t)} \widehat{\mathbf{p}}_{j}^{(t)} + \mathbf{v}^{(t)} - \mathbf{p}_{j}^{(t)}
\right\rVert_2.
\label{eq:pampjpe}
\end{equation}
To quantify skeletal plausibility, a \ac{BLL} is additionally reported over  a set of joint connections $\mathcal{E}$, and it is computed as:
\begin{equation}
\mathrm{BLL}
= \frac{1}{N_\mathrm{f}|\mathcal{E}|}\sum_{t=1}^{N_\mathrm{f}}\sum_{(i,j)\in\mathcal{E}}
\Bigl|
\left\lVert \widehat{\mathbf{p}}_{i}^{(t)}-\widehat{\mathbf{p}}_{j}^{(t)}\right\rVert_2
-
\left\lVert \mathbf{p}_{i}^{(t)}-\mathbf{p}_{j}^{(t)}\right\rVert_2
\Bigr|,
\label{eq:bone_loss}
\end{equation}
which measures the mean absolute discrepancy between predicted and reference bone lengths~\cite{sun2017compositional}. 
\ac{PA-MPJPE} measures the average joint-position estimation error after similarity-based geometric alignment, whereas \ac{BLL} evaluates the average joint-connection (i.e., bone) length estimation error.
The reported metrics are computed on the validation dataset constituted by $2$ walking rounds out of the $9$ available (round $10$ is discarded because of an incomplete motion-capture keypoint). The remaining data are used for training.

\begin{figure}[t!]
  \centering
  
    \subfloat[Validation loss\label{a}]{%
       \includegraphics[width=0.335\linewidth]{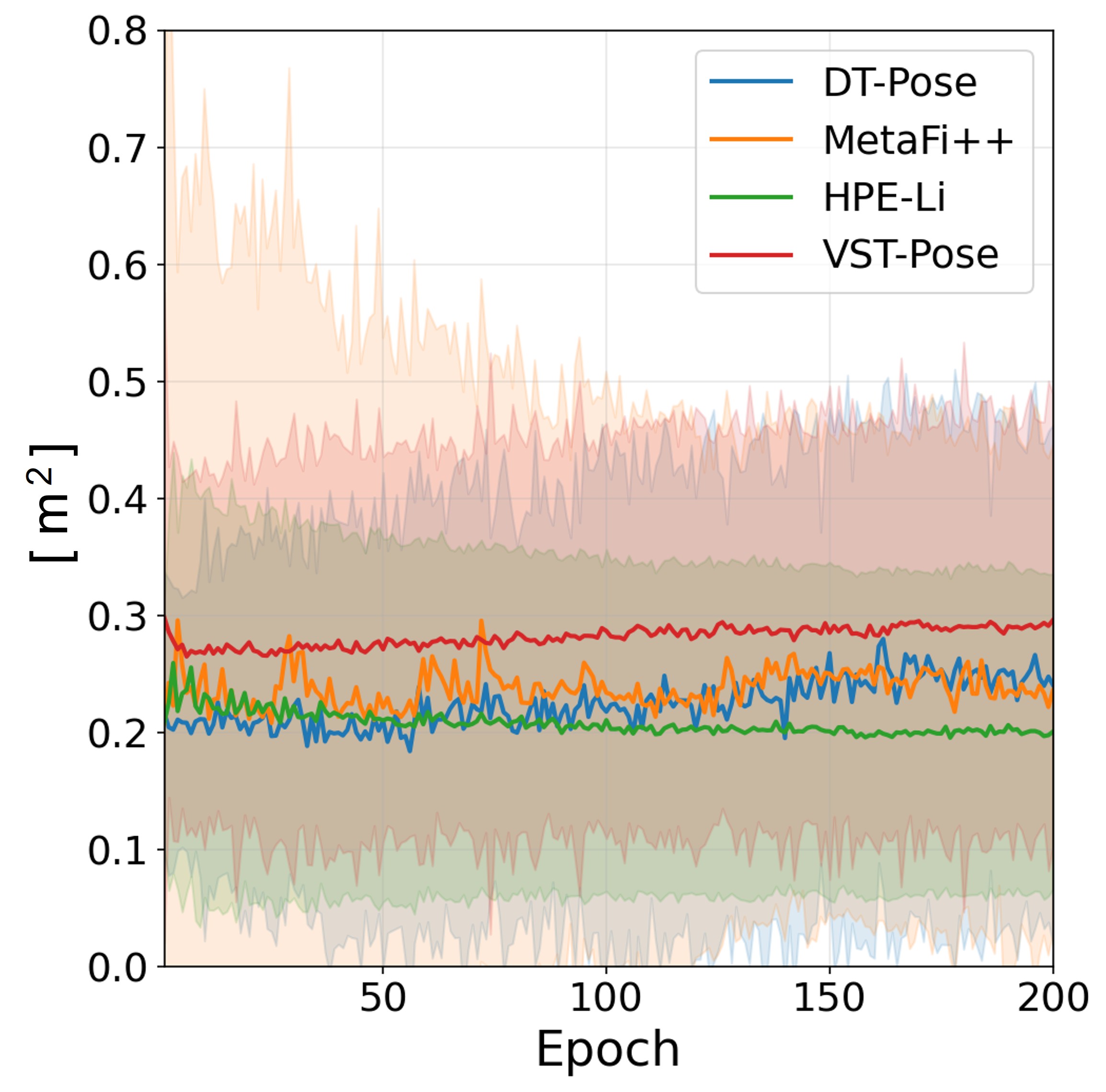}}
      \subfloat[PA-MPJPE\label{d}]{%
    \includegraphics[width=0.33\linewidth]{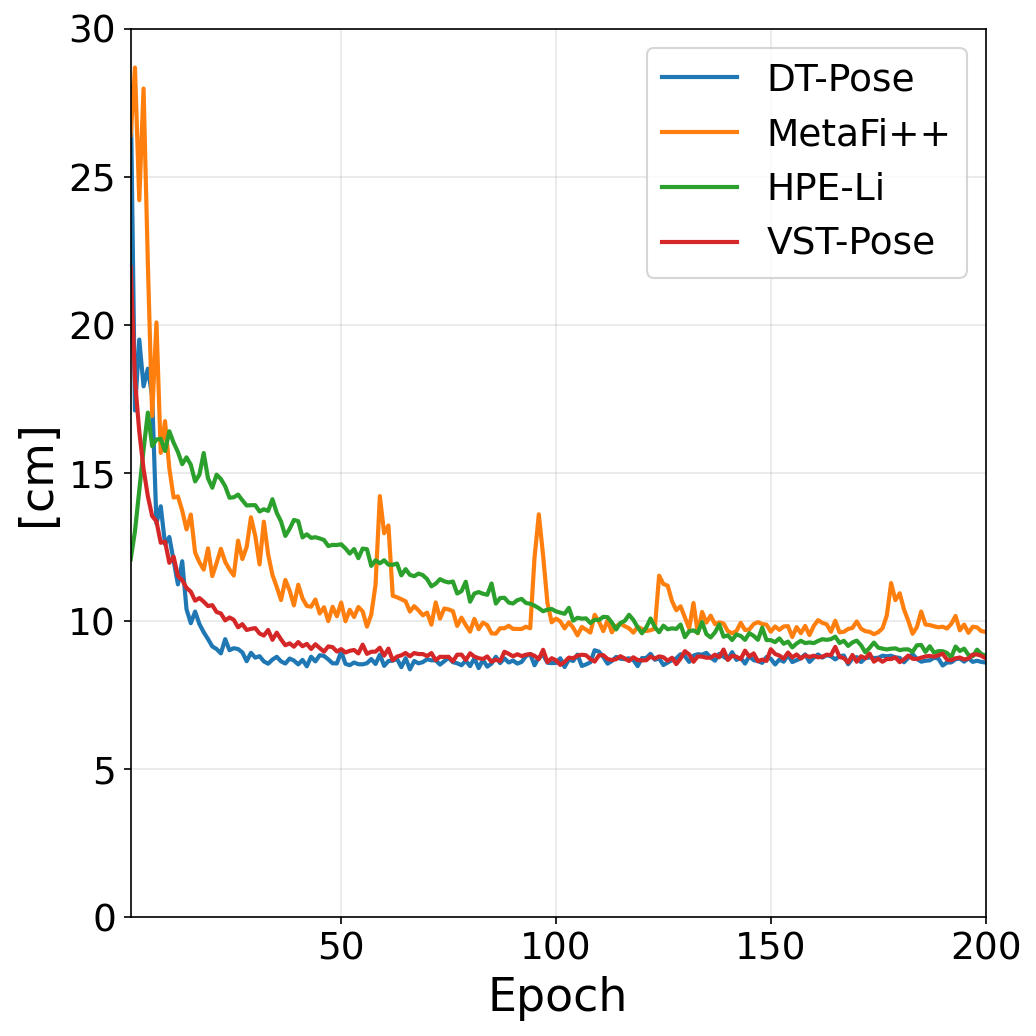}}  
    \subfloat[BLL\label{e}]{%
        \includegraphics[width=0.33\linewidth]{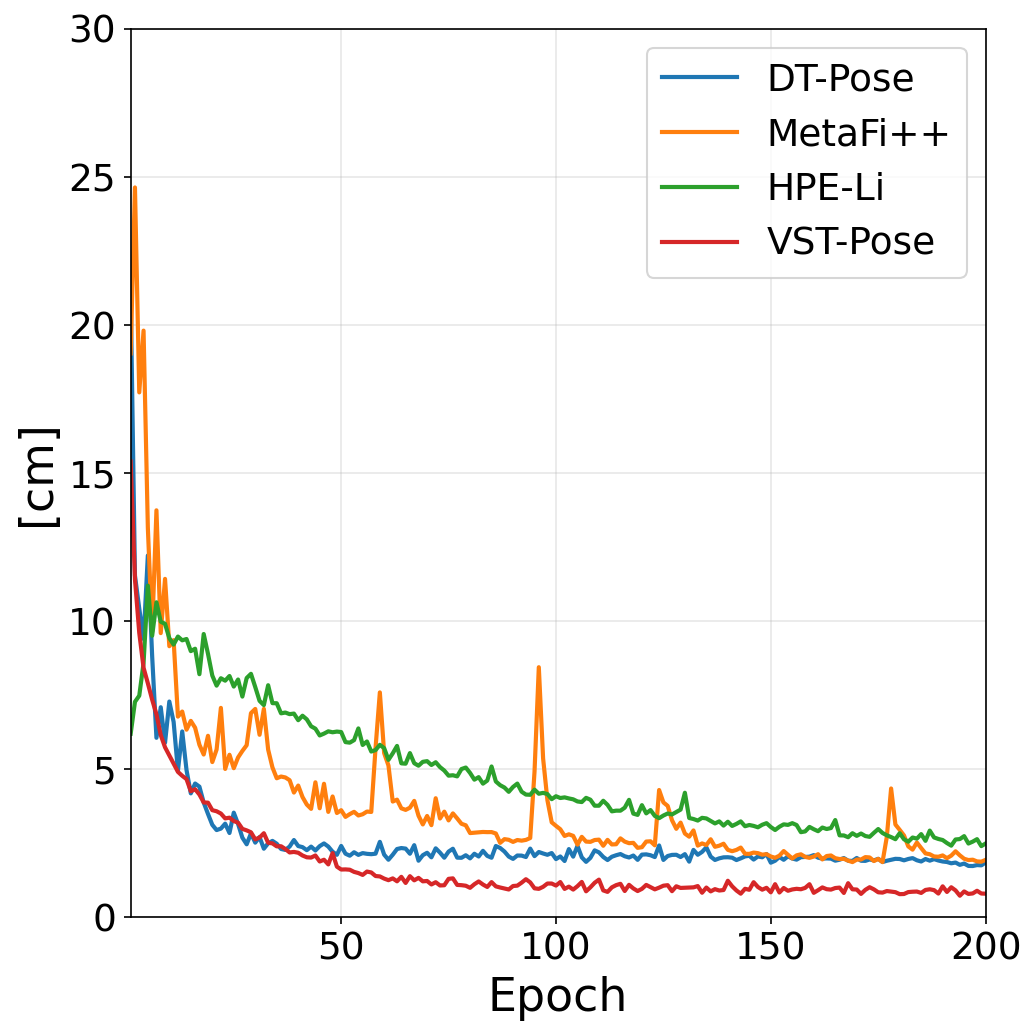}}
  \caption{
  Validation loss, \ac{PA-MPJPE} and \ac{BLL} metrics  over $200$ validation epochs, with \ac{DL} models using both amplitude and phase \ac{CSI}. 
  The shaded regions in (a) indicate the standard deviation.
  }

  \label{fig:screenshot1}
\end{figure}

\subsection{Numerical Results}
Fig. \ref{fig:screenshot1} shows the validation loss curve and the evaluation metrics over $200$ epochs. 
The validation loss remains approximately in the range of $0.2$--$0.3$\,m$^2$, with visible fluctuations across epochs, indicating that absolute coordinate regression remains challenging.
Regarding the \ac{PA-MPJPE} and \ac{BLL}, all models converge rapidly to under $10$\,cm and $3$\,cm, respectively.
A benchmarking of the four algorithms on the performance metrics is reported in Table~\ref{tab:merged_results}, where the comparison refers to the 
best model (selected by taking lowest validation \ac{MSE}).  
It reports the performance under joint amplitude-phase, amplitude-only, and phase-only CSI representations. The metrics in Table~\ref{tab:merged_results} are evaluated on the original $8$ measured keypoints, while the corresponding results for the full $13$-keypoints are reported in parentheses.
Considering the original keypoints, amplitude-only DT-Pose achieves the lowest \ac{PA-MPJPE} of $6.4$\,cm. Joint amplitude-phase input gives competitive \ac{PA-MPJPE} for HPE-Li and VST-Pose, and reduces \ac{BLL} for all models except for MetaFi++.

Analyzing the results across all models, it appears that MetaFi++ is able to provide better results with the phase-only \ac{CSI}. In contrast, the other models indicate that phase information
is more effective as a complementary feature to amplitude than
as a standalone input representation.




\begin{table}[t]
\centering
\caption{
Benchmark of best models under different CSI input based on the 8 original measured keypoints (joint-amplitude-phase / \underline{amplitude-only} / 
\textit{\scriptsize{PHASE-ONLY})}.
\footnotesize
Values inside parentheses are computed including also the extrapolated keypoints. 
}
\label{tab:merged_results}
\setlength{\tabcolsep}{4pt}
\renewcommand{\arraystretch}{1.15}
\begin{tabular}{lccc}
\toprule
\textbf{Method} 
& \textbf{Val. MSE [m$^2$]} $\downarrow$
& \textbf{PA-MPJPE [cm]} $\downarrow$ 
& \textbf{BLL [cm]} $\downarrow$ \\
\midrule
DT-Pose    
& 0.21 / \underline{0.18} / \textit{0.21}
& \begin{tabular}[c]{@{}c@{}}
7.2 / \underline{\textbf{6.4}} / \textit{16.4} \\
(9.0 / \underline{\textbf{8.0}} / \textit{18.6})
\end{tabular}
& \begin{tabular}[c]{@{}c@{}}
\textbf{1.9} / \underline{2.5} / \textit{13.7} \\
(\textbf{1.7} / \underline{2.2} / \textit{12.6})
\end{tabular}
\\
\hline

MetaFi++  
& 0.21 / \underline{0.17} / \textit{0.22}
& \begin{tabular}[c]{@{}c@{}}
9.8 / \underline{7.8} / \textit{7.5} \\
(11.6 / \underline{9.8} / \textit{9.4})
\end{tabular}
& \begin{tabular}[c]{@{}c@{}}
4.5 / \underline{2.3} / \textit{\textbf{1.9}} \\
(4.7 / \underline{2.3} / \textit{\textbf{1.7}})
\end{tabular}
\\
\hline

HPE-Li     
& 0.20 / \underline{0.17} / \textit{0.20}
& \begin{tabular}[c]{@{}c@{}}
7.0 / \underline{8.4} / \textit{11.5} \\
(9.0 / \underline{10.0} / \textit{12.8})
\end{tabular}
& \begin{tabular}[c]{@{}c@{}}
2.2 / \underline{4.3} / \textit{5.6} \\
(2.1 / \underline{4.2} / \textit{5.6})
\end{tabular}
\\
\hline

VST-Pose   
& 0.27 / \underline{0.27} / \textit{0.32}
& \begin{tabular}[c]{@{}c@{}}
6.9 / \underline{8.3} / \textit{11.1} \\
(8.9 / \underline{9.8} / \textit{13.0})
\end{tabular}
& \begin{tabular}[c]{@{}c@{}}
\textbf{1.9} / \underline{3.4} / \textit{6.2} \\
({1.9} / \underline{3.4} / \textit{6.1})
\end{tabular}
\\
\bottomrule
\end{tabular}
\vspace{-8pt}
\end{table}

\begin{figure}[t] 
\centering
   \subfloat[DT-Pose\label{5a}]{%
       \includegraphics[width=0.27\columnwidth, trim=0 0 0 0, clip]{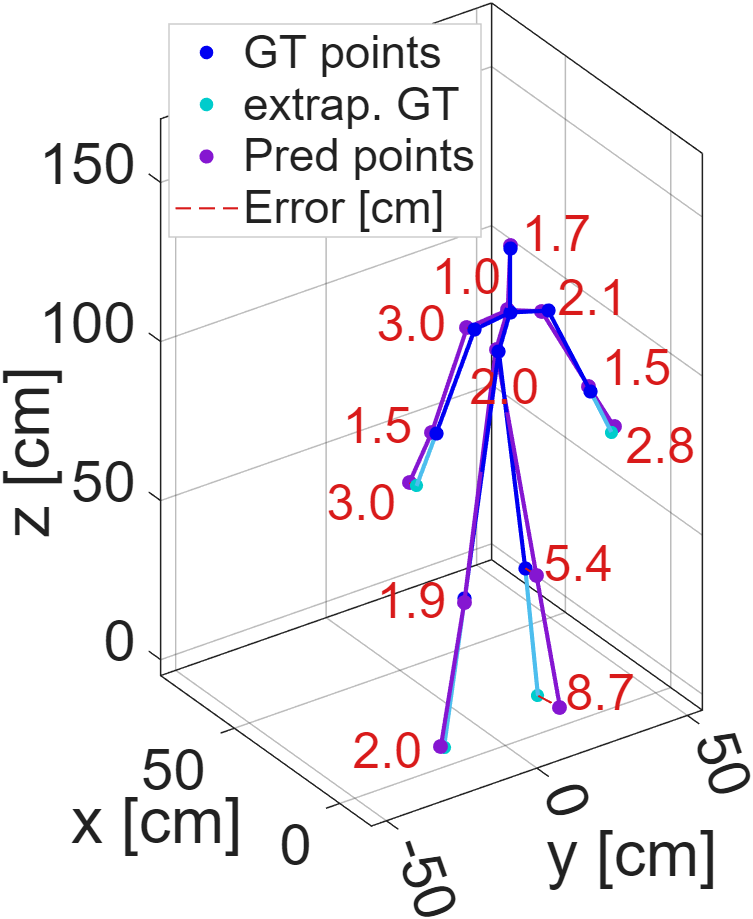}}
  \subfloat[$t=10$ \label{5b}]{%
       \includegraphics[width=0.37\columnwidth, trim=0 0 0 0, clip]{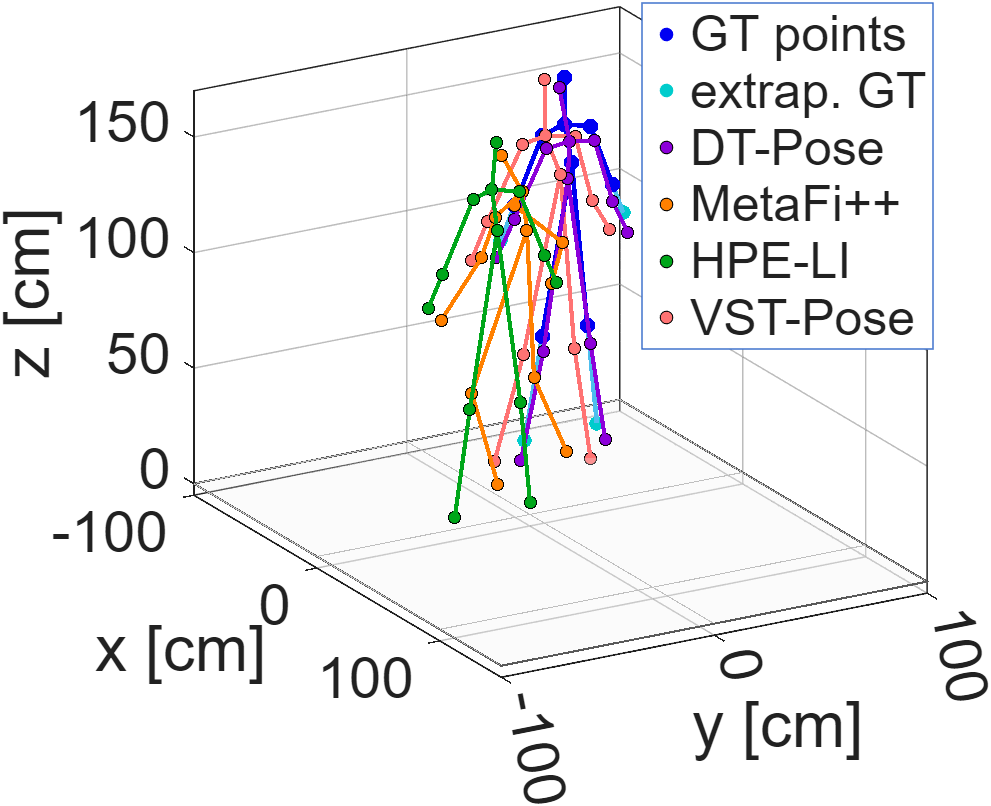}}
  \subfloat[$t=300$\label{5c}]{%
       \includegraphics[width=0.37\columnwidth, 
       trim=0 0 0 0, clip]{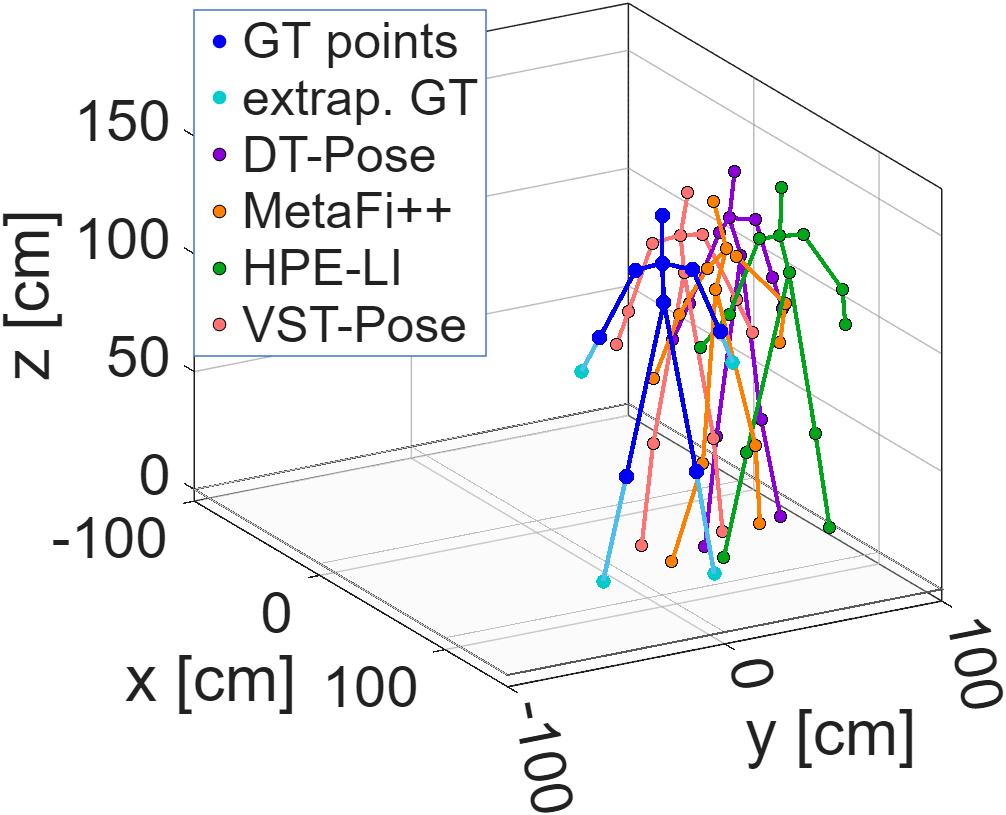}}
\caption{(a) Human pose reconstruction of DT-Pose after Procrustes alignment at $t=10$.  Raw keypoint outputs of the models: (b) $t=10$, (c) $t=300$.}
\label{fig:rmse_comparison} 
\end{figure}

Fig.~\ref{5a} illustrates the Procrustes aligned keypoint predictions at time frame $t=10$ obtained by the best DT-Pose model using both amplitude and phase. For the same time frame, Fig.~\ref{5b} reports the raw (i.e., without PA) keypoint outputs for the four models. The figure highlights that the reconstructed human skeleton is visually plausible, even if the estimate is biased with respect to the ground truth. Such error (around $50$\,cm) is also visible in Fig.~\ref{5c}, which corresponds to $t=300$, confirming that the models mainly recover relative pose configurations rather than accurate absolute positions.

\section{Conclusion}
We studied indoor human walking pose estimation from 6\,GHz \ac{CSI} in a multistatic setup.
Four state-of-the-art \ac{CSI}-to-pose deep learning models, namely MetaFi++, HPE-Li, DT-Pose, and VST-Pose, were adapted to process \ac{OFDM} \ac{CSI} and evaluated using amplitude-only, phase-only, and joint amplitude-phase representations. 
The results show that passive wireless sensing provides reliable human pose reconstruction, estimating plausible human skeletons. Even though the average error on the absolute keypoint location is around $50$\,cm, the overall skeleton is reliably reconstructed.
Overall, amplitude information carries most of the discriminative content required for pose estimation, while phase measurements are more effective as a complementary feature than as a standalone input. Phase-only \ac{CSI} appears advantageous only  for MetaFi++.
These findings suggest that \ac{CSI}-based pose estimation methods can be successfully extended from conventional Wi-Fi sensing to \ac{OFDM}-based sensing systems operating in the 6~GHz band.





\section*{Acknowledgment}
We would like to thank the authors of~\cite{usrpofdm} for the dataset.

\balance
\bibliographystyle{IEEEtran}
\bibliography{refs} 

\end{document}